\documentstyle[12pt]{article}
\input epsf.tex

\newcommand{\be}{\begin{equation}}
\newcommand{\ee}{\end{equation}}

\date{}
\title{
{\large\rm DAMTP-97-102}\hfill\vspace*{0cm}\\
{\large\rm October 1997}\hfill\vspace*{2.5cm}\\
Gauge Invariance and Factorisation\\ in Exclusive Meson Production
\\[2cm]}
\author{
A  Hebecker and P  V  Landshoff\\
{\normalsize\it DAMTP, Cambridge University, Cambridge CB3 9EW, England}
\vspace*{3cm}\\
}
\begin{document}

\setlength{\baselineskip}{18pt}
\maketitle
\begin{abstract}
\noindent
The structure of the nonperturbative vector meson vertex function 
complicates the proof of the factorisation theorem for the reaction 
$\gamma ^*p\to Vp$. It leads to additional contributions but, in a simple 
model for the vertex function, gauge invariance ensures that they cancel and 
factorisation is preserved. 
\end{abstract}
\thispagestyle{empty}
\newpage

\subsubsection*{1. Introduction}
\begin{figure}[ht]
\begin{center}
\vspace*{-.5cm}
\parbox[b]{15.7cm}{\epsfxsize=15.7cm\epsfbox{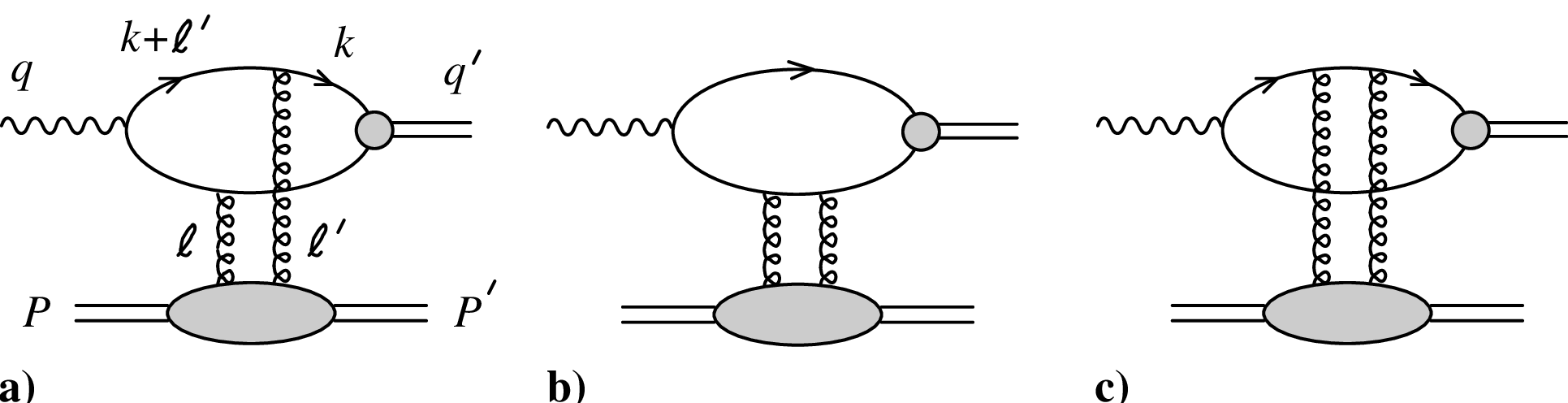}}\\
\end{center}
\refstepcounter{figure}
\label{lo}
{\bf Fig.\ref{lo}} 
The leading amplitude for a structureless meson vertex.
\end{figure}

With the measurement of exclusive vector meson electroproduction at high 
photon energy and virtuality at HERA \cite{exp} a new possibility for 
investigating the interplay of hard and soft physics is emerging. 
The original crude calculations of the process \cite{ad}, based on 
the diagrams of figure~\ref{lo}, have since been refined by several 
authors \cite{pc}. These diagrams contain a hard part, calculated 
with perturbation theory, but there is also a nonperturbative part. 
Recently, the mechanism of factorisation of the calculable hard amplitude 
and the nonperturbative component, given by the non-diagonal parton 
densities and the meson wave function, has been discussed in a rather 
general framework \cite{cfs}. 

It is the purpose of the present paper to  discuss the problems associated
with the structure of the vertex function at the top right-hand corner
of the diagrams. This couples the vector meson to the internal quark
loop. It is nonperturbative and so our knowledge of it is far from complete, 
though its analytic structure is known \cite{elop}. If we neglect spin, it 
is a function $V(u,v)$ of the squared 4-momenta on the two quark legs,
$u=k^2$ and $v=(q'-k)^2$. In the case of a nonrelativistic system, such
as the $\Upsilon$, it is a good approximation to take $V$ to be
strongly peaked at values of $u$ and $v$
such that the quarks are very close to their mass shells. But it is 
doubtful \cite{hls} that this is a good approximation for the $J/\psi$, and
for the $\rho$ surely it is not. In this case, $V(u,v)$ has branch points in
each of its variables, $u$ and $v$, associated with ``normal thresholds''
and possibly also ``anomalous thresholds'' \cite {elop}.
We show that even in leading power
this results in a breakdown of the desired factorisation  of the sum
of the diagrams of figure~\ref{lo}. 

However, this does not necessarily imply that the factorisation theorem is 
invalid. The set of diagrams of figure~\ref{lo} is not QCD gauge invariant. 
To achieve gauge invariance, they must be supplemented by other graphs in
which either or both of the gluons couples directly into the 
nonperturbative vector meson vertex function. The question we discuss
is whether these additional diagrams restore the factorisation theorem.

We are not able to give a definitive answer to this question, because
of the need to introduce two further nonperturbative vertex functions,
and we know as little about these as we do about $V$. However, we report
a calculation based on an explicit simple model in which we find that
the factorisation theorem is indeed restored. This encourages the belief
that its validity may be general.

Our model is a simple one in which all three vertex functions have the 
analytic structure that is expected on general grounds \cite{elop} and in 
which they are related in such a way that the complete leading-power 
amplitude for the exclusive meson production is gauge invariant. We then 
find that the amplitude contains leading contributions from diagrams that 
cannot be separated into a hard production process and soft wave function 
corrections. However, gauge invariance leads to a cancellation 
of the unwanted contributions and leaves the factorisation theorem intact.
The final formula can be obtained from the diagrams of figure~\ref{lo} by a 
redefinition of the rules for calculating them: basically, this amounts to 
ignoring the branch points in the vertex function $V$.

In section 2, we review some aspects of the factorisation theorem
in the case where the structure of the nonperturbative vertex function $V$
may be neglected. In section 3 we explain the complications that arise when 
its structure is taken into account, and introduce a simple model for this 
structure. Its consequences are  analysed in section 4.

\subsubsection*{2. Structureless vertex function}
According to usual ideas, unless \cite{hls} the vector meson is at least as 
heavy as the $\Upsilon$, its complete vertex function $V(u,v)$ 
is nonperturbative for small values of $u$ and $v$. Its high-$u$ or
high-$v$ tail may be calculated from perturbative QCD, but we do not consider
this tail because it contributes a correction to the amplitude 
that is a nonleading power of the perturbative coupling $\alpha _S(Q^2)$.
To begin with, we take the nonperturbative vertex to be structureless.
For simplicity, we take the meson to be spinless, and also pretend that
the photon and the quarks have no spin. 
The coupling of the gluons to the scalar quarks is given by $-ig\,(r_\mu+
r_\mu')$, where $r$ and $r'$ are the momenta of the directed quark lines,
and the coupling of the photon is $ie$, where $e$ has the dimensions
of mass. 
We use light-cone co-ordinates and work in a frame where $q,q'$ 
and $P,P'$ have zero transverse components and large `$+$' and 
`$-$' components respectively. 
We  concentrate on the forward production, so that
the transverse component of 
the momentum transfer $\Delta=q'-q=\ell-\ell'=P-P'$ vanishes, 
$\Delta_\perp =0$. 

For scalar quarks, there is a seagull two-gluon vertex. In order to obtain a 
gauge-invariant set of diagrams, we must supplement those of figure~\ref{lo} 
with diagrams that involve the seagull. However, we shall calculate in 
Feynman gauge and suppose that $W^2\gg Q^2$, where $W$ is the $\gamma ^*p$ 
energy, $W^2=(q+P)^2$. In this case the diagrams that involve the seagull 
contribute to the amplitude a nonleading power of $W$: the diagrams of 
figure~\ref{lo} are the only ones that contribute in leading power when $V$ 
is constant. We study the case of large $Q^2$, $Q^2\gg m_V^2$, so that we 
neglect all masses. Then for forward production 
$\Delta =x_{\mbox{\scriptsize Bj}}\,P$. 

The lower bubble in the diagrams of figure~\ref{lo} in principle has a 
complicated structure. However, we shall
assume that the main contribution comes from values of its subenergy 
$\sigma$ that are not too large,
$\sigma =(P-\ell)^2\ll W^2$. Then $\ell _+\ll q_+$ and the dependence of 
the lower bubble on $\sigma$ may be approximated by $\delta (\sigma )$. 
Further, when the upper parts of the diagrams are added together the
main contribution arises from small values of
$\ell_-,\quad\!\!\ell _-\ll P_-$, so that also $\ell '_-\ll P_-$ 
and $\ell ^2\sim\ell^{\prime 2}\sim\ell_\perp^2$. This may be seen most 
simply by calculating the imaginary part of the amplitude, where the left-%
most of the two quarks to which $\ell'$ is attached is on shell.
Hence the important part of the lower bubble effectively has the structure
\be
F^{\mu\nu}(\ell,\ell',P)\approx\delta(P_-\ell_+)\,F(\ell_\perp^2)\,P^\mu 
P^\nu\,,\label{fd}
\ee
which is defined to include both gluon propagators and all colour factors. 
A similar expression was found by Cheng and Wu \cite{cw} in
a tree model for the lower
bubble, though we do not need to restrict ourselves to such a simple model.
We assume that $F$ restricts the gluon momentum to be soft, 
$\ell_\perp^2\ll Q^2$. In the high energy limit it suffices to calculate 
\be
M=\int\frac{d^4\ell}{(2\pi)^4}T^{\mu\nu}F_{\mu\nu}\approx\int
\frac{d^4\ell}{4(2\pi)^4}T_{++}F_{--}\,,\label{it}
\ee
where 
\be
T^{\mu\nu}=T^{\mu\nu}(\ell,\ell',q)=T^{\mu\nu}_a+T^{\mu\nu}_b+T^{\mu\nu}_c
\ee
is the sum of the upper parts of the diagrams in figure~\ref{lo}. 

\def\half{{\scriptsize\frac{1}{2}}}
The lower amplitude $F_{\mu\nu}$ in the diagrams of figure 1 is symmetric with
respect to the two gluon lines. This symmetry of the lower amplitude allows
us to replace the properly-symmetrised upper amplitude 
$T_{\mbox{\scriptsize sym}}^{\mu\nu}$ with the unsymmetrised amplitude
$T^{\mu\nu}$ corresponding to the sum of the diagrams in figure 1. The
symmetrised amplitude is
\be
T_{\mbox{\scriptsize sym}}^{\mu\nu}(\ell,\ell',q)=\half [
T^{\mu\nu}(\ell,\ell',q)+ T^{\nu\mu}(-\ell',-\ell,q)]\,.
\ee
The two exchanged gluons together must form a colour singlet and so, at 
least for our calculation that takes account only of the lowest order in 
$\alpha_S(Q^2)$, the symmetrised
amplitude $T_{\mbox{\scriptsize sym}}^{\mu\nu}$ 
satisfies the same Ward identity as for two photons:
\be
T_{\mbox{\scriptsize sym}}^{\mu\nu}(\ell,\ell',q)\ell_\mu \ell_\nu'=0\,.
\label{w1}
\ee
Writing this equation in light-cone components and setting $\ell_\perp=
\ell_\perp'$, we see that for the small values of $\ell _-$, $\ell '_-$,
$\ell_+$ and $\ell_+'$ that we need,
\be
T_{\mbox{\scriptsize sym,}++} \sim\ell_\perp^2\,
\ee
for $\ell_\perp^2\to 0$. Here we have used the fact that the tensor 
$T_{\mbox{\scriptsize sym}}^{\mu\nu}$, which is built from $\ell'$, $\ell$ 
and $q$, has no large `$-$' components. The $\ell_-$ integration makes this 
equation hold also for the original, unsymmetrised amplitude: 
\be
\int d\ell_-T_{++} \sim\ell_\perp^2\,.\label{w2}
\ee
This is the crucial feature of the two-gluon amplitude that will simplify 
the calculation and lead to the factorising result of the next sections. 

It is convenient to begin with the contribution from diagram a) of 
figure~\ref{lo} to the $\ell_-$ integral of $T_{++}$, which is required in 
(\ref{it}):
\be
\int d\ell_-T_{a,++}=-4eg^2q_+\int\frac{d^4k}{(2\pi)^3}\,\frac{z(1-z)}
{N^2+(k_\perp+\ell_\perp)^2}\,\frac{V(k^2,(q'-k)^2)}{k^2(q'-k)^2}\,.
\label{ta}
\ee
Here $N^2=z(1-z)Q^2$, $z=k_+/q_+$ and the condition $\ell_+=0$, enforced 
by the $\delta$-function in (\ref{fd}), has been anticipated. 

Now $\int d\ell_-T_{b,++}$ and $\int d\ell_-T_{c,++}$ each carry no 
$\ell_\perp$ dependence. So to ensure the validity of (\ref{w2}) the sum of 
the three diagrams must be 
\be
\int d\ell_-T_{++}=4eg^2q_+\int\frac{d^4k}{(2\pi)^3}z(1-z){\cal N}
\frac{V(k^2,(q'-k)^2)}{k^2(q'-k)^2}\,,
\label{intit}
\ee
where
\be
{\cal N}=\Big [\frac{1}{(N^2+k_\perp^2)^2}-\frac{1}
{N^2+(k_\perp+\ell_\perp)^2}\Big ]\sim \frac{\ell_\perp^2}{N^4}\,.
\label{intitt}
\ee
We have used the softness of the wave function, which results in the 
dominant contribution to the integral arising from values of 
$k_\perp^2\ll Q^2$, and the rotational symmetry, which makes 
$k_\perp\cdot\ell_\perp$ integrate to 0. Note the $1/Q^4$ behaviour obtained 
after a cancellation of $1/Q^2$ contributions from the individual diagrams. 
This cancellation, which is closely related to the well-known effect of 
colour transparency \cite{ct}, has been discussed in \cite{ad} in the 
framework of vector meson electroproduction. 

Introduce the light-cone wave function of the meson
\be
\psi(z,k_\perp^2)=-\frac{iq'_+}{2}\int dk_-dk_+\,
\frac{V(k^2,(q'-k)^2)}{(2\pi)^4
k^2(q'-k)^2}\,\delta (k_+-zq'_+).\label{wv}
\ee
The final result following from (\ref{it}) and (\ref{intit}) is a 
convolution of the production amplitude of two on-shell quarks and the 
light-cone wave function:
\be
M=ieg^2W^2\left(\int\frac{d^2\ell_\perp}{2(2\pi)^3}\ell_\perp^2
F(\ell_\perp^2)\right)\int dz\int d^2k_\perp\frac{z(1-z)}{N^4}
\psi(z,k_\perp^2)\,.\label{lot}
\ee
This corresponds to the $O(\ell_\perp^2)$ term in the Taylor expansion
of the contribution (\ref{ta}) from figure~\ref{lo}a).

In a more general analysis the above $\ell_\perp$ integral of $F$ has to be 
replaced by an expression proportional to the non-diagonal gluon density 
which introduces an $W^2$ dependence and an additional $Q^2$ dependence.

\subsubsection*{3. Modelling the meson wave function}
In the previous section we have supposed that the vertex function $V$ 
has no structure. We used the gauge invariance of the amplitude to 
argue that there is a cancellation among the three diagrams of 
figure~\ref{lo}a). Of course, we may instead obtain the same result by 
explicit calculation of each of the three diagrams. This may be done most 
simply by calculating their imaginary parts and using the known fact that 
the complete amplitude must be pure-imaginary when its energy dependence is 
$(W^2)^{1.0}$. Alternatively, the contributions to the amplitude itself may 
be calculated by doing the $k_-$ integration, completing the integration 
contour with an infinite semicircle and taking appropriate pole residues. 

Consider figure~\ref{lo}b) for example. In either method of calculation,
if $V$ has no structure the upper quark line gets put on shell.
However, if we now take account of the known structure of $V$, there is
an additional contribution to the imaginary part corresponding to cutting
the graph through the vertex function --- the $k_-$ integration would
need to take account also of branch points of $V$, not just the poles
of the propagators. The gauge-invariance argument breaks down because
the set of diagrams by itself is no longer gauge invariant: one must
add to it diagrams where either or both of the gluons couples directly
into the vertex function.

\begin{figure}[ht]
\begin{center}
\vspace*{-.5cm}
\parbox[b]{13.1cm}{\epsfxsize=13.1cm\epsfbox{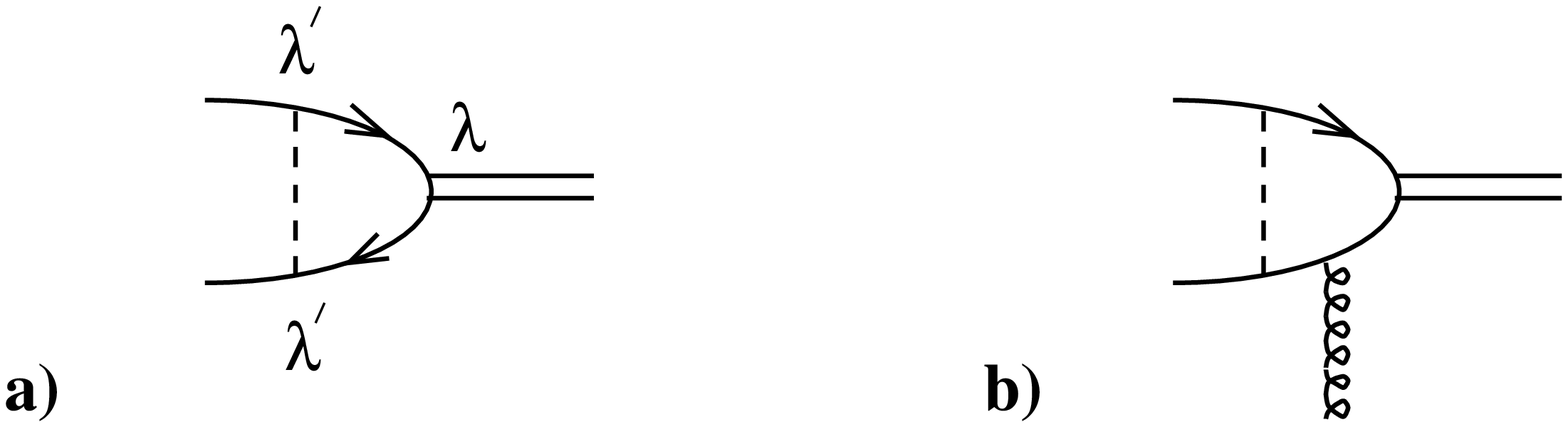}}\\
\end{center}
\refstepcounter{figure}
\label{ver}
{\bf Fig.\ref{ver}a)} Model for the nonperturbative vertex function $V$.
{\bf b)} Diagram with a gluon coupling into the vertex.
\end{figure}

In order to study this, we use the simplest model for the quark-quark-meson 
vertex function $V(u,v)$ that incorporates at least part of its known 
branch-point structure. It is the purely nonperturbative vertex function 
that we are modelling: it goes to zero suitably rapidly when either of the 
squared 4-momenta $u$ or $v$ of the quarks becomes large. We do not consider 
its perturbative tail, which would be obtained by exchanging a perturbative 
gluon between the quarks. It is a familiar notion \cite{elop} that the 
correct {\it analytic} properties of nonperturbative amplitudes are those 
corresponding to Feynman graphs, even though the {\it numerical} values of 
such graphs have no physical significance. In order to model the vertex 
function $V$, therefore, we use the simple Feynman graph of 
figure~\ref{ver}a), where the line that joins the quarks is a scalar (like 
the quarks themselves in our simple calculations) which couples to them with 
strength $\lambda'$ and where the right-hand internal vertex is taken to be 
a constant $\lambda$. This model has branch points in each of the variables 
$u$ and $v$, and it has the appropriate softness. 

\begin{figure}[ht]
\begin{center}
\vspace*{-.5cm}
\parbox[b]{5.7cm}{\epsfxsize=5.7cm\epsfbox{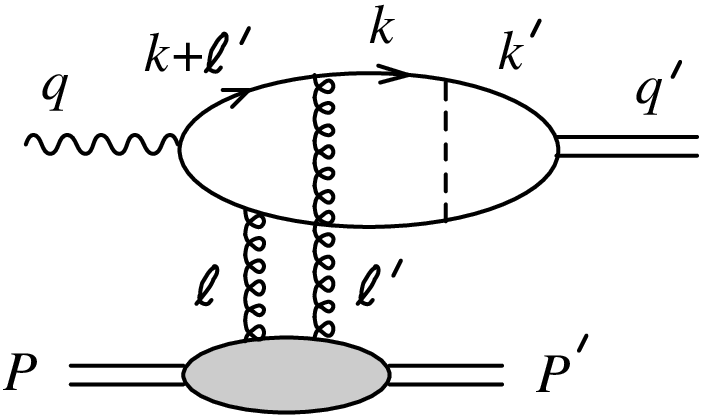}}\\
\end{center}
\refstepcounter{figure}
\label{nlo}
{\bf Fig.\ref{nlo}} 
Diagram for meson production with the vertex modelled by scalar 
particle exchange.
\end{figure}

When the vertex function of figure~\ref{ver}a) is used in figure~\ref{lo}a), 
we obtain figure~\ref{nlo}. The expression for the upper part of the diagram 
is (\ref{ta}) with 
\be
V(k^2,(q'-k)^2)=\int\frac{d^4k'}{(2\pi)^4}\,\frac{i\lambda\lambda'^2}{k'^2
(q'-k')^2(k-k')^2}\,.\label{tf2v}
\ee
The diagram of figure~\ref{nlo} by itself gives no consistent description of 
meson production since it lacks gauge invariance. This problem is not cured 
by just adding the two diagrams \ref{lo}b) and c) with the blob replaced by 
the vertex of figure~\ref{ver}a). It is necessary also to include 
diagrams where a gluon is coupled into the vertex (see, e.g., 
figure~\ref{ver}b)). Furthermore, diagrams where both gluons couple into 
the vertex have to be included. 

As will be demonstrated in the next section, the complete result can 
nevertheless be extracted directly from (\ref{ta}) and (\ref{tf2v}).

\subsubsection*{4. Gauge invariance and factorisation}
\begin{figure}[ht]
\begin{center}
\vspace*{-.5cm}
\parbox[b]{15.7cm}{\epsfxsize=15.7cm\epsfbox{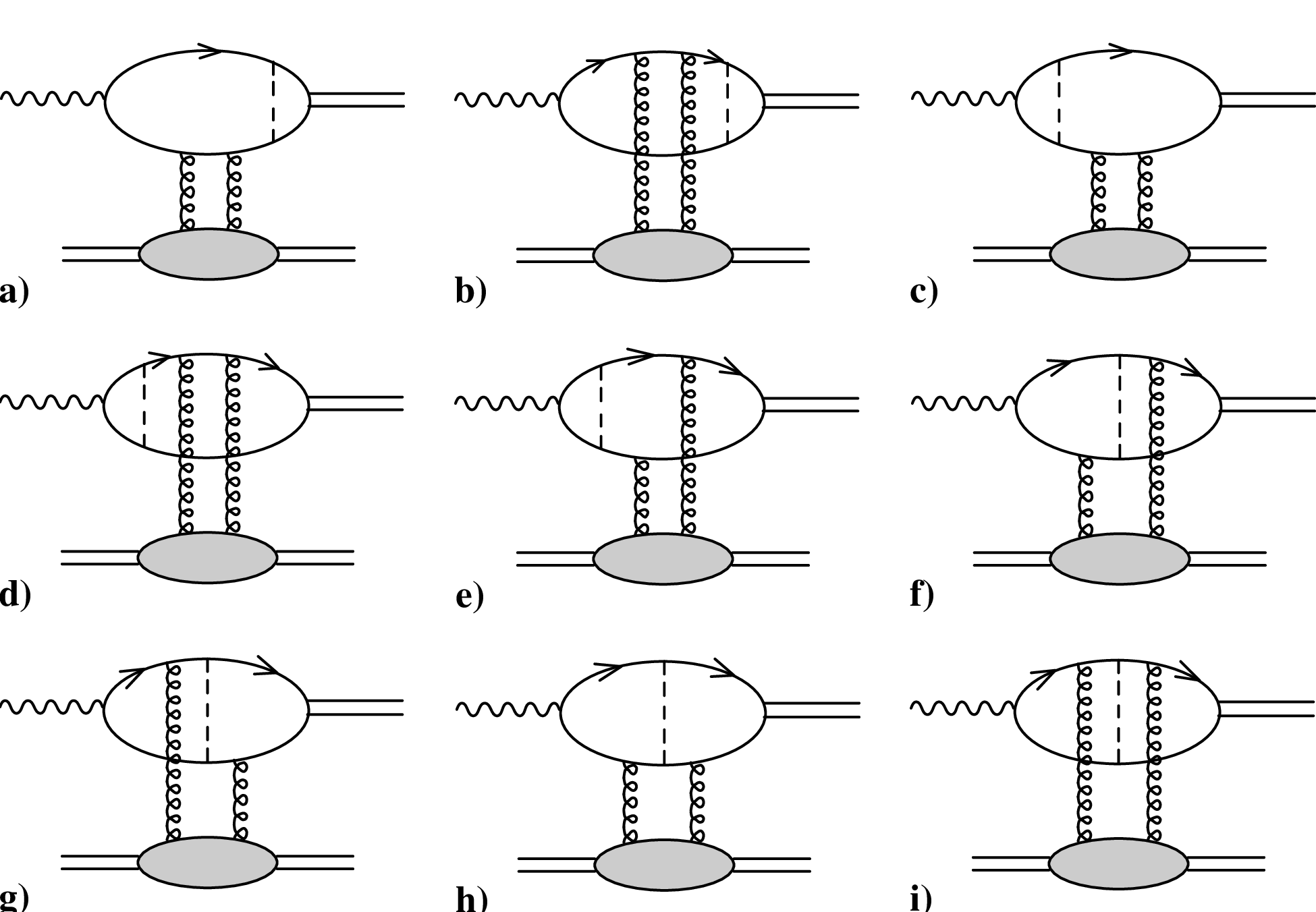}}\\
\end{center}
\refstepcounter{figure}
\label{rest}
{\bf Fig.\ref{rest}} 
The remaining diagrams contributing to meson production within the above 
simple model for the meson wave function.
\end{figure}

To obtain gauge invariance within the model of the vertex illustrated in 
figure~\ref{ver}a), diagrams with the two gluons attached to the quark loop 
in all possible ways have to be added to that of figure~\ref{nlo}. In 
addition to figure~\ref{nlo} there are nine such diagrams. They are shown in 
figure~\ref{rest}. 

The same gauge invariance arguments that lead to (\ref{w2}) apply to 
the sum of all the diagrams in figures~\ref{nlo} and \ref{rest}. Therefore, 
the complete result for $T_{++}$, which is now defined by the sum of the 
upper parts of all these diagrams, can be obtained by extracting the 
$\ell_\perp^2$ term at leading order in $W^2$ and $Q^2$. Such a term, with 
a power behaviour $\sim \ell_\perp^2W^2/Q^4$, is obtained from the diagram 
in figure~\ref{nlo} (see (\ref{ta}) and (\ref{tf2v})) by expanding around 
$\ell_\perp=0$. We now argue that none of the other diagrams 
gives rise to such a leading-order $\ell_\perp^2$ contribution. 

It is simpler to evaluate the contributions to the amplitude itself, 
rather than its imaginary part.
We continue the evaluation of the diagram in figure~\ref{nlo} by performing 
the $k_-$  and $k_-'$ integrations in (\ref{ta}) and (\ref{tf2v}). 
Consider the region where $k_+>k_+'$ and pick up the poles in such a way 
that the propagators with momenta $k'$ and $q'-k$ go on-shell. The 
resulting expression reads 
\begin{eqnarray}
\int d\ell_-T_{\mbox{\scriptsize figure~\ref{nlo}},\,++}&=&\frac{-ieg^2
\lambda\lambda'^2q_+}{(2\pi)^5}\int dzd^2k_\perp\frac{z(1-z)}{[N^2+(k_\perp+
\ell_\perp)^2]k_\perp^2}\times\label{expl}
\\ \nonumber\\
&&\int dz'd^2k_\perp'\frac{1}{[(z-z')(k_\perp'^2/z'+k_\perp^2/(1-z))+
(k_\perp-k_\perp')^2]k_\perp'^2}\,,\nonumber
\end{eqnarray}
where $z'=k_+'/q_+$ and the integration is restricted to $z>z'$. A 
corresponding expression can be obtained for the region $z<z'$.

The above manipulations can also be applied to all the diagrams of 
figure~\ref{rest}. One arrives at the following results. The upper parts of 
diagrams a) -- d) have no $\ell_\perp$ dependence at all since the 
$\ell_-$ integration puts the quark propagator connecting the two gluon 
vertices on-shell. Diagrams h) and i) vanish after the $\ell_-$ integration 
because the integrand has two poles, both of which lie on the same side of 
the real axis. To analyse the remaining diagrams e) -- g) it is convenient 
to route the momentum $\ell$ as far as possible to the left. After all the 
`$-$' integrations are performed it becomes obvious that each diagram 
contributes only at order $W^2/Q^4$. To obtain an $\ell_\perp^2$ term one of 
the off-shell propagators left of the gluon vertices has to be expanded 
around $\ell_\perp=0$. This brings the power down to $W^2/Q^6$. Therefore, 
no leading-order $\ell_\perp^2$-contribution is obtained. 

The above discussion shows that the complete answer is given by the 
$\ell_\perp^2$ term from the Taylor expansion of (\ref{ta}). 
The amplitude $M$ is 
precisely the one of (\ref{lot}) and (\ref{wv}), with 
$V(k^2,(q'-k)^2)$ given by (\ref{tf2v}).

We have also checked the correctness of this simple factorising result by 
explicitly reducing all diagrams of figure~\ref{rest} to a form 
similar to that given 
in (\ref{expl}) for the diagram of figure~\ref{nlo}. The required 
cancellations occur on the level of the integrands, before the $k_\perp$, 
$k_\perp'$, $z$ and $z'$ integrations.

\subsubsection*{5. Conclusions}
The mechanism of factorisation in exclusive meson production has been 
analysed in the framework of a simple scalar model. In this model the meson 
is formed by two scalar quarks interacting via the exchange of a scalar 
boson. From a calculation of
all  contributing diagrams within the restriction of two-%
gluon exchange the following picture emerges.

The complete result contains leading contributions from diagrams that 
cannot be factorised into quark-pair production and meson formation. 
Nevertheless, in Feynman gauge the answer to the calculation
can be anticipated by looking only at one 
particular factorising diagram. The reason for this simplification is gauge 
invariance. In the dominant region where the transverse momentum $\ell_\perp$ 
of the two $t$-channel gluons is small,  gauge invariance requires the 
complete quark part of the amplitude to be proportional to $\ell_\perp^2$. 
The leading $\ell_\perp^2$ dependence comes exclusively from one 
diagram. Thus, the complete answer can be obtained from this particular 
diagram, which has the property to factorise explicitly if the two quark 
lines are cut. The resulting amplitude can be written in a factorised form. 

It was our intention to demonstrate the above mechanism of gauge invariance 
induced cancellations in as simple and as explicit a way as possible.
Our analysis supplements the
otherwise much more general and complete discussion of \cite{cfs} 
by making it more explicit and by handling the known structure of
the nonperturbative meson vertex function. 

Our calculation has many important limitations. It is restricted to the very 
simplest model of the meson vertex function, with only the simplest features 
of the known branch-point structure. However, a generalisation to a more 
complicated wave function structure seems straightforward. We have assumed 
that the main contribution to the diagrams in figures~\ref{lo}, \ref{nlo} 
and \ref{rest} comes from values of the subenergy of the lower bubble that 
are not too large. Nevertheless, the integration over this subenergy will 
provide the energy growth of our cross section, corresponding to the rapid 
rise at small $x$ of the gluon distribution observed at HERA. We have not 
discussed the subtleties involved in factorising the off-diagonal gluon 
density, nor the behaviour in the region where the longitudinal momentum 
fraction of one of the quarks becomes small.

\bigskip\bigskip
{\sl We would like to thank John Collins and Markus Diehl for helpful 
discussions and comments.

This research was supported in part by the UK Particle Physics and
Astronomy Research Council and by the EC Programme
``Training and Mobility of Researchers", Network
``Hadronic Physics with High Energy Electromagnetic Probes",
contract} ERB FMRX-CT96-0008.

\end{document}